# How Cyber Criminal Use Social Engineering To Target Organizations.

Shazny Ahmed Hussain Nismy
*Department of Computing and Informatics*
Bournemouth University
Bournemouth United Kingdom
S5063123@bournemouth.ac.uk

Amrut Mahesh Kajave
*Department of Computing and Informatics*
Bournemouth University
Bournemouth United Kingdom
s5431775@bournemouth.ac.uk

*Abstract* — Social engineering is described as the art of manipulation. Cybercriminal use manipulation to victims their targets using psychological principles to change their behavior to make unconscious decisions. This study identifies the attack and techniques used by cybercriminal to conduct social engineering attacks within an organization. This study evaluate how social engineering attacks are delivered, techniques used and highlights how attackers take advantage Compromised systems. Lastly this study will also evaluate and provide the best solutions to help mitigate social engineering attacks with an organization.

## I. INTRODUCTION

The practise of manipulating others to gain access to or accomplish a goal is known as social engineering (Mouton et al. 2016).In other words, social engineering refers to the application of psychology to the newest technologies, such as using phishing emails or smishing, to attack victims. Most social engineering assaults have origins in cyber security attacks, and the majority of the top cyber security attacks began with social engineering(gov 2021). Social engineering uses a variety of tactics and techniques that have an impact on both individual users and the digital world. In this paper, we have mainly discussed how social engineering affects both businesses and individual users. Following an introduction to social engineering and its various forms, the delivery methods and strategies for social engineering are discussed, followed by the problems of social engineering. The relationship between psychological principles and social engineering has now been clarified. In the end, research was conducted, and a suggestion was made.

## II. SOCIAL ENGINEERING

Human trust one another which result in the disclosure of personal information. Cybercriminals are aware of this and will use social engineering technique manipulate victim for selfish gain (Junger et al. 2017) Social engineering is described as the art of manipulation. As time goes on social engineering attack are forever increasing, resulting in millions of users being victimized by social engineering attacks (Salahdine and Kaabouch 2019). Cybercriminal use multiple attack vector to exploit their victims such a phishing, Smishing, vishing and many other attack vectors which will be discussed further in the study. Furthermore, when conducting these social engineering attacks, the cybercriminal uses the attack vector to exploit the behavior of the targeted individual by changing their behavior using psychological factor. The psychological factor that cybercriminal exploit on the individual are Authority, intimidation, consensus, scarcity, familiarity, trust, and urgency on the (Siddiqi et al. 2022). exploiting the psychological factors of a targeted Results in the victim making unconscious decision which - exploits targeted victims' cognitive abilities to make decisions (Rodriguez et al. 2022). In this study will critically evaluate how social engineering attack take places and how the attack exploit organization. furthermore, it will also look at which solution can be put at an early stage to help mitigate social engineering attack with in are organization. moreover, it vital the organization are a aver of the attack vector which attackers use to conduct social engineering attack.

## III. TYPES OF SOCIAL ENGINEERING ATTACKS

Social engineering is the process of using deception or persuasion to get knowledge that may have negative effects. Social engineering can be roughly divided into two categories in the actual world: human-based social engineering and technological-based social engineering. No technology is used in a human-based attack; all operations are done through psychological manipulation or secret information listing. When obtaining information in a technical context, a variety of technologies and approaches are used. There are several social engineering attacks causes huge loss to the industries such as (Department of Justice 2019) In year 2019, phishing fraud against Google and Facebook by a fake company that pretends to be legitimate by provide invoices and setting up the bank account with a business name. The attacker sent invoices to employees to show Attentively. The social engineering attacks which are all time popular are Phishing, Vishing and Smishing which still exists today. Also, other modern types such as water holing, baiting, tailgating and many more have roots of Phishing, Vishing and Smishing.

### A. Phishing

Phishing is primarily a method used to fraudulently acquire sensitive data about a business or a particular person. The success rate is still astonishingly high despite being the oldest tactic ever utilized by attackers (Sam 2022). Phishing is sending potential victims emails that appear legitimate but are attacks. Additionally, in the modern era, victims have encountered click bait, which is used on some websites to entice visitors to participate in activities like giving out personal and financial information to phony websites that pretend to have offers (Berkenkopf 2016).Through a well-known job board organization named CareerBuilder, many targets were subjected to a more sophisticated phishing attack. Attackers pose as regular job applicants in this case, but instead of including a resume, they upload dangerous files such as malware. These resumes were then forwarded by CareerBuilder to several employers. The most successful attack witnessed the spread of malware to numerous companies (Jai 2015). Another well-known attack was the IRS refund, in which hackers used official employees by sending them emails about tax refunds that appeared legitimate. When the recipients read the emails, they download documents that contain malware and end up infecting the company's computers. This hack potentially steals firm data and result in losses (Thorne and Stryker 2015)

*B. Smishing*

Smishing is a mobile phone assault that uses the Short Message Service (SMS). It is an SMS-based variation of phishing. It starts with an SMS message pointing the victim to a website that can serve up numerous attack vectors, including malware. This attack succeeds mostly because of the message's use of urgency and intimidation, which may contain a warning like "Your credit score is crucial, go here to view," along with a URL that requests personal and financial information that is sufficient to compromise the target (Mishra and Soni 2019).There are numerous instances where smishing affects businesses. In the case of Uber, a teenager used this method to gain access to internal emails, cloud storage, and source code repositories. He did this by sending a phishing message to an employee of the company that claimed it required "two-factor authentication," and then using that information to compromise the data. Then, Uber handled the problem by portraying the occurrence as part of its bug bounty program and rewarding the teenager (Niall 2022).

*C. Vishing*

Vishing, also known as voice over phishing, is a form of phone-based phishing where the attacker utilizes a voice call instead. This type of advanced phishing assault involves the attacker making a fraudulent call while posing as a representative of a particular institution, such as a bank, the public sector, or another entity that is interested in sensitive information. The majority of phishing attacks take place in groups where attackers work together to plan successful attacks. In order to pull off this fraud, the attacker group in India teaches minors how to speak in English and financial industry jargon. In India, this scam is seen as having a high level of threat (NIHARIKA 2021). This technique is typically used as an extension of email phishing to get a target to provide sensitive information. Normally, a toll-free number is provided, and when the victim calls it, the rogue interactive voice response system is reached. The system will encourage the target to divulge some verification data. It is typical for the system to reject input from a target in order to make sure that several PINs are revealed. This is sufficient for the assailant to continue and steal money from a target, whether it be a person or an organization.

## IV. SOCIAL ENGINEERING DELIVERY METHODS

Social engineering attack are classed as the art of manipulation. To conduct these attacks cybercriminal will need to target their victim/organization with delivery method. The delivery method commonly used by cyber criminals are Spam over Email, Spam over Instant Messaging, Spear Phishing, and Whaling (Sabillon et al. 2016; Nurse 2019). Using a reliable social engineering delivery method will allow the attacker to conduct a precise attack for the indented purpose to compromises a targeted organization. The delivery method will breach CIA of the compromised system. Data gather from (statista 2022) show the 66 % of organizations faced at least 1 spear phishing attacks within 2020. Furthermore, research conducted by(Jyothiikaa 2022) has also highlighted the 73 % of spam emails are Phishing attacks.

*A. Spam*

Spam is unwanted mail or junk mail which is sent to multiple recipients. Spamming a phishing attack can target a large amount of user within a organization with same established Phishing email. Highly established company send spam emails to organization for advertisement for sales. However cyber criminals send spam email as mail contains the link to the phishing sites or with hidden malicious software (Malware) to steal confidential data within the target system(Patil et al. 2017). Any type of spam Phishing attack is to gain sensitive information such as usernames, passwords, and credit card details commonly an attack for financial gain with a malicious reason(Mandadi et al. 2022). The spam email masquerades themselves as a trustworthy email. Furthermore, it has also been highlighted the 70% of phishing emails are opened by their targets and that 90% of security breaches in companies are a result of phishing attack (Nikolina 2022) .

*B. Spam over Instant Messaging (SPIM)*

Mobile phone usage has increased enormously over the year. research conducted by (Jack 2022) has highlighted that there will be around 7.52 billion smartphone users by 2026. The high volume of smartphone users is resulting in a major challenge for Instant Messaging. As cybercriminal are targeting employees within organization with SPIM which contains malicious software. Researcher has highlighted that limited attention has been given to the detection of SPIM as compared to email SPAM(Maroof 2010). However, an enormous amount of research has highlighted that SPIM is far more dangerous as it hard to detect compared to email SPAM. Furthermore, Instant messages is more sociable and more user friendly. As a victim can just click on the URL, the worms can be downloaded, and the devices can be infected and compromised (Maroof 2010; Das et al. 2012).

*C. Spear Phishing*

Spear phishing is a phishing attack. However cyber-criminal conduct Spear phishing attack on a specific individual. The aim is to target a specific victim and compromises his/her confidential data to get access into a sensitive system where the attacker can install malware to steal confidential data (Atmojo et al. 2021). Spear phishing attacks are executed by sending malware or sending a URL link to the targets using fake e-mail address to manipulate the victim to believe they are receiving a trusted email, Cybercriminal know that employees are the weakest links in security that they can put an organization at risk of a cyber-attack. by targeting employees with various social engineering techniques they are able manipulate their target to provide sensitive data (Shakela and Jazri 2019; Atmojo et al. 2021) .

## V. SOCIAL ENGINEERING TECHNIQUES

The main purpose of conducting social engineering attacks is to gather sensitive information from the targeted victim. The information gathered will be vital data for the reconnaissance stage of the cyber-attack targeting organization. The main objective within the reconnaissance stage is to gather physical or virtual information (Ahmad et al. 2015)The information gathered will play a vital part towards breaching the organization confidentiality, integrity, and availability (CIA). To conduct a successful attack the criminal will use social engineering techniques to manipulate the targeted victims. The techniques used by the attacker for social engineering attacks are Dumpster Diving, Shoulder Surfing, Prepending, Pharming, Tailgating, Eliciting Information, Watering Hole Attack. (Tajpour et al. 2013; Ahmad et al. 2015; Mike and David 2021). In this study we will be looking at each of following social engineering techniques mentioned above and analyzing the techniques in more detail on how the attackers are targeting the organization and manipulating their victims using the social engineering techniques. The techniques will be discussed below in more detail.

*A. Dumpster Diving*

One of the Physical ways of gathering data involves researching in the trach of a targeted victim or organization to find sensitive information such as credentials (Username and password), filenames, or other pieces of confidential data which can support towards compromising a critical system or user account (Koyun and al Janabi 2017). Dumpster diving attacks occur within an organization due to the lack of policy and procedure in places towards disposing sensitive information. As large number of employees are known to dispose confidential data in clear text as well as disposing hard drive and computer technology which contain sensitive information. (Cazier and Weaver 2007; Kirk et al. 2019).

*B. Shoulder Surfing*

Another form of Physical social engineering attack is Shoulder Surfing it involves directly looking over someone's shoulder or using a magnification from farther distances to observe a user activity to gather sensitive information (Kirk et al. 2019). This type of social engineering attack can be done in multiple different ways within an organization. However, Shoulder Surfing attacks target user who are accessing or inputting sensitive information in a manner that allows direct observation (Mbuguah and Otibine 2022). Nevertheless, Shoulder Surfing Attacks can be conducted from a distance or even using surveillance technology this allow the attacker to gather data on schedule or routine activities of the victim to compromise a critical system and breach the CIA.

*C. Typosquatting*

Typosquatting also known as URL hijacking is the most common way cyber criminals conduct social engineering attacks and manipulate user. The attack is conducted based on typographic errors that are hard to notice while quickly reading the URL of the website. employees within organization are manipulated to believe URLs created with Typosquatting to be trusted domain which manipulate them clicking a link to an untrusted website (Aldawood and Skinner 2019a). An example of Typosquatting is www.g00gle.com where the cybercriminal has impersonated google by Replacing O with a zero (0). Typosquatting is the most popular way cybercriminal get employees within organization to visit extremely dangerous website (Buber et al. 2017).

*D. Prepending*

Prepending is a Techniques used for social engineering attacks. The attacker will add something to the beginning of a statement to manipulate a victim. Within the social engineering context, the attackers are known to victimize employees within an organization by acting to supplying information that another will act upon. For example, Prepending can use the psychological concept of authority to manipulate employees to gather sensitive information by informing that the boss has sent them to retrieve the data(Tech 2022). As the attacks has informed that the information is requested by the boss the employees will be less likely to withhold the requested information by the attacker.(Williams et al. 2018).

*E. Pretexting / Reverse social engineering*

Pretexting techniques manipulate the victim into a narrative to giving up some item of information. The social engineering attack of Pretexting conducted inform of impersonation. For example, Cyber criminals would call up, masquerading as a fellow employee from sales team within the organization to gather sensitive information (Zhang et al. 2012). The pretext information provided by the attacker does not have to be true it just needs to convince to victim to provide help. This attack can target organization in multiple way such as Phishing, Smishing and Vishing(Greitzer et al. 2014).

*F. Pharming*

The most effective way to deliver a phishing attack is using the Pharming Technique. the Pharming Technique is like a phishing attack as the victim will believe that they are entering their personal credentials such as (username, password) to a legitimate website. however, the attacker is targeting the victim to a spoofed website where the inserted information such username and password get sent to the attacker (Chantler and Broadhurst 2008). Unlike the other Technique this attack is made possible by the attacker poisoning the DNS server to change the IP Address of the legitimate wed site to a spoofed Ip address which attack has full access to(Schipke 2006).

*G. Tailgating*

A common technique used by cybercriminal is Tailgating. The action of Tailgating is following an oblivious human target with valid access through a secure door into a restricted space. This attack targets the lack of access control measures which have been put in places within the organization(Breda et al. 2017). The attacker will request the victim to hold the door or can simply reach for it and enter before it closes following the victim into a restricted area (Ahmad et al. 2015). Furthermore, Attackers conducting a Tailgating attack can impersonate as a cleaner or maintenance worker to get access into the organization to be more discreet and follow employee throw secure doors.

VI. CHALLENGES IMPOSED BY SOCIAL ENGINEERING ATTACKS

Businesses need technology to function in daily life, and when technology is involved, social engineering attacks are at risk. The greatest threat to employees in the "Finance" industry comes from factors like exploiting human weakness, which is done without the use of technical skills. Attackers first learn about victims through their attitudes before launching a coordinated information attack to obtain information or steal money. Another difficulty is "Uncertainty," in which unforeseen attacks occur with no advance notice, combining social engineering with zero-day attack (Li et al. 2019). The UK government estimates that 39% of firms experienced cyberattacks in 2022, of which 83% were phishing attempts (Robbie and Maddy 2022). This statistic highlights the seriousness of social engineering. However, to counteract these problems, (Aldawood and Skinner 2019b)proposed a "training and awareness programme" that would provide training and awareness to stay up to date on all aspects related to business environment, social, constitutional, organizational, economical, and personal by addressing social engineering threats.

### A. Identity Fraud

The use of fraudulent credentials to gain access to intellectual property is known as identity fraud. Also known as using someone else's identity as a goal or primary tool while doing an illegal act (Koops and Leenes 2006). According to a press release from the Federal Trade Commission, the number of identity theft and fraud complaints in the United States in 2021 will total over $5.8 billion, with impostor scams alone costing around $2.3 billion (Jay 2022).Identity fraud and identity theft have a hazy line between them; for example, identity theft is the illegal act of taking information, while identity fraud is the use of stolen information. Additionally, strategies for reducing fraud are offered by organizations such two-factor authentication, avoiding public Wi-Fi, and others (Sumanth 2022).

### B. Invoice Scams

The purpose of phishing assaults, which sometimes include invoice scams, is to trick a firm into paying for something it has not bought. Simply issue a false invoice that appears genuine to receive payment. Things considered, despite most businesses having robust accounting systems, an outside entity may send an email invoice purporting to represent an organization's part (Morris 1996). Adani and Essar Group in India have been implicated in a power traffic fraud that is thought to be worth 50,000 crores of rupees. Scams allegedly involve coal purchase invoices and construction equipment (PARANJOY GUHA 2016). Scams involving invoices pose a serious threat, particularly to small enterprises where accounting involvement is minimal. It has been suggested to prevent and reduce invoice fraud by (Teerakanok et al. 2020).

### C. Credential Harvesting

An attacker can access a number of passes to the system by using a process known as credential harvesting, which involves gathering credential information like user-id, password, and so on. When a user clicks a link in a phishing email that looks exactly like a login page, the attacker receives the information as soon as it is submitted, and the victim is then sent to an error page with the message "page could not be found." This is a typical method of credential harvesting (Br and Shivaleela 2929). Attackers today, including Emotet, Black Basta, and Fin7, plot their attacks in groups to carry them out successfully. Cybercrime conducted by this group is described in the article (CyberWire staff 2022)

### D. Impersonation

An impersonation attack is a form of social engineering in which an attacker pretends to be a different individual or poses as a valid user (or group of users) in order to obtain data they are not authorized to have. In this kind of attack, the attacker frequently employs social engineering techniques to learn more about the system and/or target, such as impersonating an IT department employee and requesting login details (Chen et al. 2017).In COVID-19, an attacker took advantage of the victim's circumstances, leading to the discovery of more than 15000 cases by the middle of 2020 and a loss of more than £58 million(UK Finance Limited 2022).

## VII. PSYCHOLOGICAL PRINCIPLES OF SOCIAL ENGINEERING

The psychological components of manipulating human emotions against economic objectives are the focus of social engineering terminology. Curiosity, excitement, fear, or greed are the human emotions that drive people to open phishing emails and fall prey to them(Abraham and Chengalur-Smith 2010). The author tries to demonstrate how harmful email designs make use of emotive themes, like the Black Friday sale, when people are eager for excellent deals, but attackers' profit from this circumstance (James 2022). Similar to this, gambling addicts may receive messages about winning the jackpot or receiving bonuses. Out of greed, the victim may give up their bank information, which can result in financial loss (Ben and Stephanie 2022),Another instance of pop-up messages on browsers that can result in fraud and compromise your social media accounts is the "AxLocker" ransomware, which can remove the discord credentials and has a lot of examples where it has harmed people (Bill 2022).

### A. Authority

In order to convince their victim that they have the authority to handle the data or service that can manage the event, an attacker will often pretend to be a high-level official. This is known as the "authority" principle of social engineering. The victim feels that giving the attacker access to the data will solve the problem in this instance. In their study, the authors (Bullée et al. 2015). Present several cases involving the concept of authority. Additionally, they contrasted a variety of investigations and tests pertaining to the concept of authority. In the article(Ferreira and Lenzini 2015), a thorough review of the phishing attach concepts is provided.

### B. Intimidation

The intimidation principle is a stricter version of the principle of authority. It can be subtle, based on perceived power, or more overt, based on the use of communication to establish dominance. By posing as a higher authority, the attacker in this case might enter the business' decision-making process (Uebelacker and Quiel 2014). As an example, it might terrify a certain authority figure into telling them to finish a few steps before moving on. The Kronos human resources firm will experience a ransomware attack in December 2021 that will disrupt customer payroll and timesheets (Jennifer 2021)

### C. Consensus

When an attacker asserts that a behavior is typical or commonly recognized within the environment of the attack, this is referred to as consensus. This choice could be manipulated to produce the desired results(Dennis 2021). Attacks accomplish this by persuading victims to believe they can be trusted. They pull this off by making it appear as though others would have trusted the assailant or already did. Fake evaluations are prevalent on fake shopping websites, giving the victim the go-ahead to proceed.(codecademy

2022)The case study of how scammers tricked victims by posing as a legitimate auto auction company was discussed in the blog of (Lauren 2022).

*D. Scarcity*

When there is a limited supply of anything, people are naturally motivated to act swiftly to get it before it disappears. This is what is meant by scarcity. This forces the victim to act quickly so that he doesn't lose the opportunity(Costa et al. 2021).Attacker employs the time limit and the obstacle restriction as two strategies(Lenny 2019)). Attacker draws victim by inflicting time emotion in time-limited situations. In one Facebook fraud, an attacker posted about an account being deactivated to con users (Graham 2011).In obstacle restriction, the attacker wants the victim to wait and take specific steps so that they would feel legitimately protected.

*E. Familiarity*

People act in favor of those they like or feel a connection to. Building this trust through familiarity and appeal may result in misguided confidence (Rusch 2003). It signifies that one person just purchases the product or utilizes the service again, not because the product is excellent but rather because the seller is more likeable or acquainted to the individual.

*F. Trust*

Trust, as far as we are aware, is defined as "a strong belief in someone or something in terms of elements of truth, character, and strength"(Dictionary 2022). But in social engineering, trust is characterized as knowing how something will behave in a particular scenario. Here, social engineers employ the approach of familiarity first to get to know the target before offering the service to win over the target's trust. Building trust may take some time at first, but once it has been established, it becomes challenging to question an attacker's conduct (Costa et al. 2021). An actual illustration would be if a hacker wanted to attack a corporation, he or she would try to communicate by pretending to be delivering the service in a legitimate fashion while stealing data and compromising without even being noticed. The process could be extremely slow, or the corporation might not pay attention to the data leak.

*G. Urgency*

There are two types of urgency: those that target a specific individual or group, like phishing emails about last-minute sales or offers that attract victims (Fakhar 2020)A second type of attack may result from familiarity, linking, and trust in which a planned mission is carried out, the system is compromised, and the same attacker seizes an opportunity to infect the company with malware. According to a report (Jessica 2019), "scenes of urgency" are the root cause of 70% of phishing attacks.

VIII. SOLUTIONS

As highlighted above, the attacker uses a variety of social engineering techniques across a variety of sites. Belove, we have suggested ways for mitigating these attacks by providing solutions.

*A. Policy and procedure:*

Every phishing attack has a business context, such as e-commerce, travel, charity, or something else entirely, which gives the victim the chance to part with money or provide information. As a result, every company has a set of legal policies, both internal and external, that serve as evidence of the data's integrity (Conteh and Schmick 2016). It is challenging to determine whether a phishing email is genuine or not as social engineering assaults get more complex. The solution is for the firm to change its policies and inform its users of those updates. Internal policies that deal with employee identity, device rights, and other issues that deal with data and could lead to future social engineering assaults must also be addressed (Ghafir et al. 2016)

*B. Training and awareness:*

The prevention of social engineering attacks includes education and awareness campaigns. Government and business organizations are actively educating the public about phishing attacks on a variety of platforms and how to spot social engineering attacks (Ui Danc 2022). educating employees and other staff members about data regulation and policy. Data importance in the digital environment requires cautious handling during training and awareness campaigns (Dao 2022)According to a recent survey, training and awareness initiatives have protected almost 80% of the population (Adenike 2022)

*C. Vulnerabilities testing throughout email phishing:*

In the past ten years, it has been difficult to uncover phishing vulnerabilities since there has been a lack of focus on cyber security during software development, as well as a lack of awareness training and lax legislation that made phishing attacks simple to carry out. These days, phishing attacks are difficult to detect since they appear authentic and have important certifications like VeriSign that attest to the legitimacy of the websites, in addition to being related to legitimate businesses (verisign 2022). There are widely used solutions that can block email phishing utilizing AI filters that can instantly determine whether an email is legitimate or not by employing blacklist and whitelist filters (Mohammad et al. 2015).Additionally, some browser extensions provide brief summaries of whether a website is safe to visit or not (SafeToOpen 2022)

*D. Multifactor authentication:*

The system or service can increase security by using two-factor or multi-factor authentication (MFA). Multi-factor authentication therefore functions by requiring additional authentication evidence before allowing access to the service. If the attacker has access to the credentials, an authentication method like "email + password" may be exploited. OTP and app-based authentication use real-time logs to avoid attacks, and multifactor authentication functions such as "email + password authentication + OTP or app login" (Moul 2019)Additionally, when someone tries to log in using an OTP notification alert with the warning, multifactor authentication serves as security monitoring that may be used. The authors of the research (Sun et al. 2022) contrasted real-time phishing attacks with multifactor authentication.

IX. CONCLUSION

Cybercriminals are always finding alternative way to conduct social engineering attack. Organizations need to provide a robust security posture which conation network layer security and physical layer security. As social engineering attack can occur in both layers. Furthermore, they will need to stay UpToDate with security, as the security environment is always change. Lastly social engineering attacks are highlighted as the art of manipulation to help prevent employees from becoming victims' organization should consider training their stuff on how cybercriminal manipulate their victim using psychological principles. As security Technology and security policies are not sufficient to protect an organization from social engineering attacks.

ACKNOWLEDGMENT

Firstly, I would like to thank Dr. Philip Davies for his teaching on how to researcher and gather valuable information for the problem environment of social engineering. Secondly, we like to give a massive thank you Bournemouth University for providing us with the chances to study and undertake our Masters' degree to become professional researchers.